      [1999/12/01 v1.4c Il Nuovo Cimento]
\documentclass{cimento}
\usepackage{graphicx}  
\title{What is the optimum stellar rotation rate for a collapsar?}
\author{W.~H.~Lee\from{IA} }
\instlist{\inst{IA} Instituto de Astronom\'{\i}a, Universidad Nacional Aut\'{o}noma de M\'{e}xico, Apdo. Postal 70--264, Cd. Universitaria, M\'{e}xico DF 04510 }
\PACSes{
\PACSit{95.30.Lz}{ 95.30.Sf}{ 95.85.Pw}{ 97.10.Gz}{ 97.10.Kc}{ 97.60.-s}{ 98.70.Rz}}
\begin{document}

\maketitle

\begin{abstract}
We consider low angular momentum, neutrino cooled accretion 
flows onto newborn black holes in the context of the 
collapsar model for long Gamma Ray Bursts, and find a 
considerable energy release for rotation rates lower than 
those usually considered. The efficiency for the transformation 
of gravitational binding energy into radiation is maximized  
when the equatorial angular momentum $l_0\simeq 2 r_{\rm g} c$.
\end{abstract}

\section{Introduction}\label{intro}

The collapsar model for GRBs \cite{w93,mw99} has proved quite
successful for interpreting the association of long GRBs with star
forming regions and with observed SNe at low redshift. It is not
entirely clear, however, that stellar cores (or which fraction of
them) will actually lead to the production of a successful GRB at the
end of their evolution, since known mechanisms (mass loss, magnetic
fields) act to spin down the inner regions of the star in various ways
\cite{h05}. We consider here the outcome of accretion onto a newborn
black hole following core collapse in the case where the stellar
rotation rate is low by the standards usually considered in collapsar
calculations.

\section{Ballistics}\label{particles}

It is an obvious, but sometimes overlooked fact, that in Newtonian
theory, a test particle in orbit about a massive point body can be
placed in a stable circular orbit for any given value of its orbital
angular momentum. The condition gives the circularization radius as
\begin{equation}
r_{\rm c}= \frac{\ell^{2}}{GM_{*}},
\end{equation}
where $\ell$ is the specific angular momentum and $M_{*}$ is the mass
of the central body. The orbit is stable precisely because $\ell(r)$
is a monotonically increasing function of radius. A perturbation
conserving angular momentum will merely induce small, epicyclic
oscillations about the equilibrium radius $r_{\rm c}$.  

In general relativity (GR) this no longer holds, and a qualitatively
different effect appears, allowing for the presence of capture orbits
even at a finite value of $\ell$. This is the unique
``pit-in-the-potential'' feature of GR, dictating that at low enough
values of $\ell(r)$, one need not lose angular momentum in order to
fall onto the central object. In terms of stability, the equilibrium
angular momentum, $J_{\rm eq}(r)$, for test particle motion exhibits a
minimum at a particular radius, $r_{\rm ms}$, termed the {\em
marginally stable} orbit. For $r<r_{\rm ms}$, stable circular orbits
are no longer possible since $dJ_{\rm eq}(r)/dr < 0$.

\begin{figure}
\includegraphics[width=7cm]{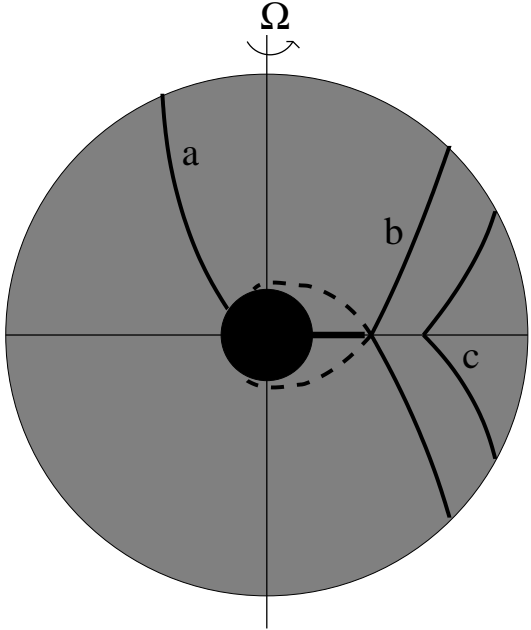}   
\caption{For a cloud rotating as a rigid body, where the specific angular 
momentum varies with the polar angle as $\ell=l_{0}\sin^{2}\theta$,
three possible types of trajectories are shown: (a) direct accretion
onto the central black hole; (b) impact with a symmetrical flowline
from the opposite hemisphere (without which direct accretion onto the
black hole would also occur); (c) encounter with the centrifugal
barrier and settling at the circularization radius in the presence of
dissipation. }
\label{flowlines}
\end{figure}

The location of the marginally stable orbit is a function of the spin
of the central object (a black hole if we are to consider a
relativistic scenario following the collapse of an iron core inside a
massive star), but is in all cases a modest number of gravitational
radii, $r_{\rm g}=2GM_{*}/c^{2}$. In particular, $r_{\rm ms}=3r_{\rm
g}; 3/2r_{\rm g}$ for a non--spinning (Schwarszchild) and maximally
spinning (Kerr) hole, respectively. Thus for this effect to have any
impact upon the evolution of infalling test masses, we must consider a
situation in which the circularization radius and that of the
marginally stable orbit are comparable, $r_{\rm c}\approx r_{\rm
ms}$. This immediately translates into the condition
\begin{equation}
\ell = f \, \times \, (r_{\rm g} c),
\end{equation}
where $f$ is a factor of order unity.

A meridional projection of flowlines for infalling particles in rigid
(slow) rotation around a black hole is shown in
Figure~\ref{flowlines}. Given the rotation law, polar inflow is
entirely radial, while along the equator the centrifugal barrier is
felt most importantly. It is evident as well that if we consider the
behavior of an accreting gas cloud, purely ballistic considerations
will not tell us the entire story, since the flow lines at small radii
and in the plane of the equator clearly cross. A shock may thus form,
and a hydrodynamical study is necessary if one wishes to proceed any
further. 

The general context here can be thought of as GRBs, but it has various
astrophysical applications, namely (and originally to our knowledge)
to the case of wind--fed neutron stars in High--Mass X-Ray Binaries
\cite{bi01}.

\section{Hydrodynamics}\label{hydro}

We have thus carried out a series of two--dimensional simulations of
low angular momentum accretion onto black holes, with some
simplifications and assumptions keeping collapsars in mind.

First, we do not perform fully GR calculations but use instead the
potential proposed by Paczy\'{n}ski \& Wiita \cite{pw80}, $\Phi_{\rm
PW}=-GM_{*}/(r-r_{\rm g})$. This reproduces the position of the
marginally stable orbit correctly for a Schwarszchild hole, albeit at
a slightly different value of the specific orbital angular
momentum. Second, we assume azimuthal symmetry in order to compute a
decent number of dynamical times in the inner regions of the flow
(several hundred). Third, we assume an initially free-falling flow in
rigid body rotation, parametrized by the equatorial angular momentum,
$l_{0}$, and the accretion rate, $\dot{M}$. Finally, we make use of
fairly detailed thermodynamics, incorporating an ideal gas of free
nucleons and $\alpha$ particles, $e^{\pm}$ pairs and photons and
neutrino cooling (due to pair captures by free nucleons and $e^{\pm}$
annihilation), while using as well the viscosity prescription of
Shakura \& Sunyaev \cite{ss73} to allow for angular momentum
transport. We refer the reader to \cite{lrr06} for further details of
the implementaion.

\begin{figure}
\includegraphics[width=\textwidth]{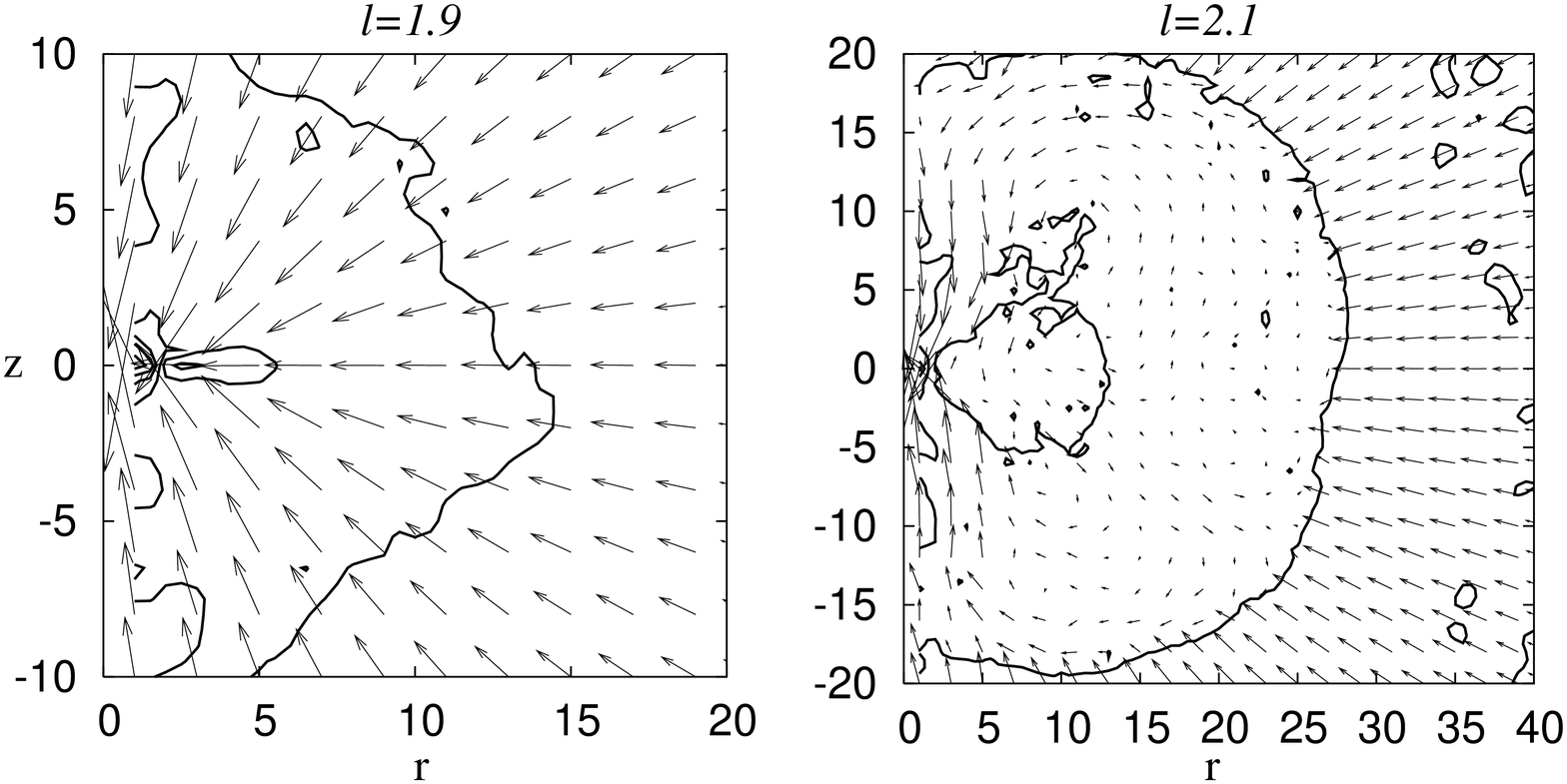}   
\caption{
Accretion flow morphologies for low (left) and high (right) specific
angular momentum calculations (in units of $r_{\rm g} c$) without the
effects of viscosity ($\alpha=0$). The dwarf disk in near free fall
and the large toroidal bubble are clearly seen in either case. The
axes are labeled in units of $r_{\rm g}$, with density contours and
the velocity field shown in each case. }
\label{types}
\end{figure}

For low angular momentum, $l_{0} < 2 r_{\rm g} c$, the resulting
configuration is completely independent of the strength of the
viscosity, and we see the formation of a dwarf disk in the equatorial
plane. This material accretes simply because it does not have enough
angular momentum to remain in orbit around the black hole (it does not
have a circularization radius). For high angular momentum $l_{0} > 2
r_{\rm g} c$, the outcome depends somewhat on the actual value of
$\alpha$, but is generically different in the sense that the dwarf
disk is replaced by a hot, shock--heated toroidal bubble that
surrounds the black hole. The qualitative nature of this transition is
independent of the assumed mass accretion rate (we explored values
between $10^{-3}$ and $0.5 M_{\odot}$~s$^{-1}$), and we show in
Figure~\ref{types} representative snapshots for the two classes of
solutions.

\section{Maximizing the power output}\label{ccl}

A common concern when addressing the issue of progenitors for long
GRBs has always been their rotation rate \cite{h05,yl05,wh06}. If the
stellar core is not endowed with sufficient angular momentum, mostly
radial inflow will ensue, and this is not a favorable situation in
terms of the potential energy release. On the other hand, a problem
that is not so frequently addressed is that if too much angular
momentum is present, the circularization radius of the gas may be too
large to produce an efficient neutrino cooled accretion flow
\cite{npk01}. A large fraction of the energy may then simply produce
outflows not directly emanating from the central region and thus not
capable of producing the required power output. The maximum power is
actually obtained by first lowering the gas as much as possible in the
gravitational potential well {\em and} simultaneously shocking it to
dissipate as much of its energy as possible, which can then be
extracted, possibly to produce a GRB. We argue here that this may
occur precisely at the transition when the flow morphology changes
from that of a hot torus to a thin, efficiently cooled dwarf disk,
when $l_{0}\approx 2 r_{\rm g} c$, or $5 \times
10^{16}$cm$^{2}$~s$^{-1}$ for a hole of $3M_{\odot}$. The efficiency is
maximized in two different ways: on one hand, the raw neutrino
luminosity, $L_{\nu}$, is highest close to this threshold, and on the
other, the most energetic neutrinos are simltaneously produced, thus
giving a relatively high efficiency for annihilation.

\acknowledgments
It is a pleasure to thank E. Ramirez-Ruiz for collaboration. Financial
support from Conacyt (45845) and DGAPA-UNAM (IN-119203) is gratefully
acknowledged.

\end{document}